\documentclass[11pt]{article}

\usepackage{amsmath}
\usepackage{amssymb}
\usepackage{graphicx}
\usepackage{xcolor}
\numberwithin{equation}{section}
\usepackage{array}
\usepackage{mathtools}
\usepackage{dsfont}
\usepackage{mathrsfs}
\usepackage{tikz}\usetikzlibrary{matrix,fit}
\usepackage{varwidth}
\usepackage{enumerate}

\usepackage{hyperref}
\newdimen\mytextwidth
\newcommand\rem[2][cyan!40!green]{\noindent\nobreak\hfil\penalty1000\hfilneg
\mytextwidth=\linewidth\advance\mytextwidth by 2mm%
\begin{tikzpicture}[baseline=-\the\dimexpr\fontdimen22\textfont2\relax]\node[outer sep=0pt,draw=black,fill=#1,fill opacity=1,text opacity=1,rectangle,rounded corners]{\begin{varwidth}{\mytextwidth}\textcolor{white}{#2}\end{varwidth}};
\end{tikzpicture}\allowbreak%
}

\newcommand{\dd}{\partial}
\newcommand{\CP}{\mathds{CP}}
\newcommand{\CC}{\mathds{C}}

\newcommand{\bea}{\begin{equation}}
\newcommand{\eea}{\end{equation}}
\newcommand{\bear}{\begin{eqnarray}}
\newcommand{\eear}{\end{eqnarray}}
\newcommand{\bearr}{\begin{eqnarray*}}
\newcommand{\eearr}{\end{eqnarray*}}

\newcommand{\appendixnumberline}[1]{Appendix.\space}

\let\oldappendix\appendix
\makeatletter
\renewcommand{\appendix}{%
  \addtocontents{toc}{\let\protect\numberline\protect\appendixnumberline}%
  \renewcommand{\@seccntformat}[1]{\large Appendix. }%
  \oldappendix
}
\makeatother

\begin{document}

\title{\vspace{-1.0cm} Ricci-flat metrics on vector bundles\\ over flag manifolds}
\author{Ismail Achmed-Zade$^{1,2}$ \;and Dmitri Bykov$^{1,2, 3}$\footnote{Emails:
I.AchmedZade@physik.uni-muenchen.de, bykov@mpp.mpg.de, bykov@mi-ras.ru}  \\ \\ 
{\small $^1$ Max-Planck-Institut f\"ur Physik, F\"ohringer Ring 6, D-80805 Munich, Germany}\\
{\small $^2$ Arnold Sommerfeld Center for Theoretical Physics,}\\ {\small Theresienstrasse 37, D-80333 Munich, Germany}\\
{\small $^3$ Steklov
Mathematical Institute of Russ. Acad. Sci.,}\\ {\small Gubkina str. 8, 119991 Moscow, Russia \;}}
\date{}

\begin{flushright}    
  {\small
    MPP-2019-81 \\
    LMU-ASC 17/19
  }
\end{flushright}

{\let\newpage\relax\maketitle}

\maketitle

\vspace{-1cm}
\begin{center}
\line(1,0){400}
\end{center}
\vspace{-0.3cm}
\textbf{Abstract.} We construct explicit complete Ricci-flat metrics on the total spaces of certain vector bundles over flag manifolds of the group $SU(n)$, for all K\"ahler classes. These metrics are natural generalizations of the metrics of Candelas-de la Ossa on the conifold, Pando Zayas-Tseytlin on the canonical bundle over $\CP^1\times \CP^1$,  as well as the metrics on canonical bundles over flag manifolds, recently constructed by  van Coevering. 
\vspace{-0.9cm}
\begin{center}
\line(1,0){400}
\end{center}

\tableofcontents

\section{Introduction and main result}

The problem of constructing explicit Ricci-flat K\"ahler metrics is rather complicated. In the compact case no such metrics are known, due to the fact that the condition of Ricci-flatness implies the absence of non-parallel Killing vectors~\cite[Section~1.84]{Besse}. However, in the non-compact case symmetries may be present, and the metrics are sometimes known in explicit form. The early examples include the Eguchi-Hanson `gravitational instanton'~\cite{EH} and its generalizations by Gibbons and Hawking~\cite{GH}. Another example that will be important for us is the metric of Candelas-de la Ossa~\cite{CdO} on the so-called `resolved conifold' and its immediate generalization constructed in~\cite{PZTmain}. Various other metrics are known: those of cohomogeneity one of Stenzel~\cite{Stenzel1993} and Nitta~\cite{Nitta:2003yk} as well as the higher-cohomegeneity metrics on manifolds that admit Killing-Yano tensors~\cite{Gauduchon, LuPope1}. One can also construct hyperk\"ahler metrics on the cotangent bundle of flag manifolds using the hyperk\"ahler quotient construction of~\cite{nakajima1994}. In the simplest case of the Grassmannians this was elaborated upon in \cite{BiquardGauduchon1996}. For the purposes of the present paper, however, it will be sufficient to understand in detail the examples of~\cite{EH, CdO, PZTmain}.

An important feature of the metric of \cite{EH} is that it may be thought of as the metric on the total space of the canonical bundle $\mathcal{O}(-2)$ over $\CP^1$. 
In fact, it is a simple application of the so-called Calabi ansatz~\cite{Calabi79}, which allows constructing a complete Ricci-flat metric on the canonical bundle of a K\"ahler-Einstein manifold $\mathcal{M}_{KE}$ of positive curvature -- in this case this manifold is simply $\CP^1$ with its Fubini-Study (round) metric.

An interesting generalization may be obtained by replacing the base manifold $\CP^1$ by a manifold of flags in $\CC^n$. We recall that a flag manifold may be specified by a sequence of increasing integers $0<m_1<\ldots <m_s=n$ which define the flag
\bea\label{flag1}
0\subset L_1\subset \ldots \subset L_s=\CC^n\,,
\eea
where $L_k$ are linear subspaces of $\CC^n$, such that $\mathrm{dim}\,{L_k}=m_k$. In place of $m_k$ we will often use the integers $n_i$, defined by $m_k=\sum\limits_{i=1}^k\,n_i$. In these terms, the manifold of flags~(\ref{flag1}) is a homogeneous space
\bea\label{flag2}
\mathscr{F}_{n_1,\ldots, n_s}={U(n)\over U(n_1)\times ... \times U(n_s)}\,.
\eea
In what follows we will sometimes use the short notation $\mathscr{F}$ for this manifold. The complex geometry of flag manifolds was first studied in the classical work~\cite{BorelHirzebruch58}.

The authors of~\cite{CorreaGrama17} use Calabi's ansatz to construct a Ricci-flat metric on the canonical bundle of the flag manifold, equipped with such a K\"ahler-Einstein metric. An important characteristic of Calabi's ansatz, that we review in Section~\ref{cp1cp1}, is that it produces a metric with a \emph{fixed} K\"ahler class. On the other hand, by the Calabi-Yau theorem~\cite{Calabi, Yau1, Yau2}, in the case of a compact manifold with $\mathbf{c}_1=0$ there should exist a Ricci-flat metric in \emph{every} K\"ahler class. The relevant non-compact generalization of the theorem (to asymptotically-conical spaces), with the same statement, was constructed in~\cite{Coevering, Goto}.

\begin{figure}[h]
    \centering
    \includegraphics[width=\textwidth]{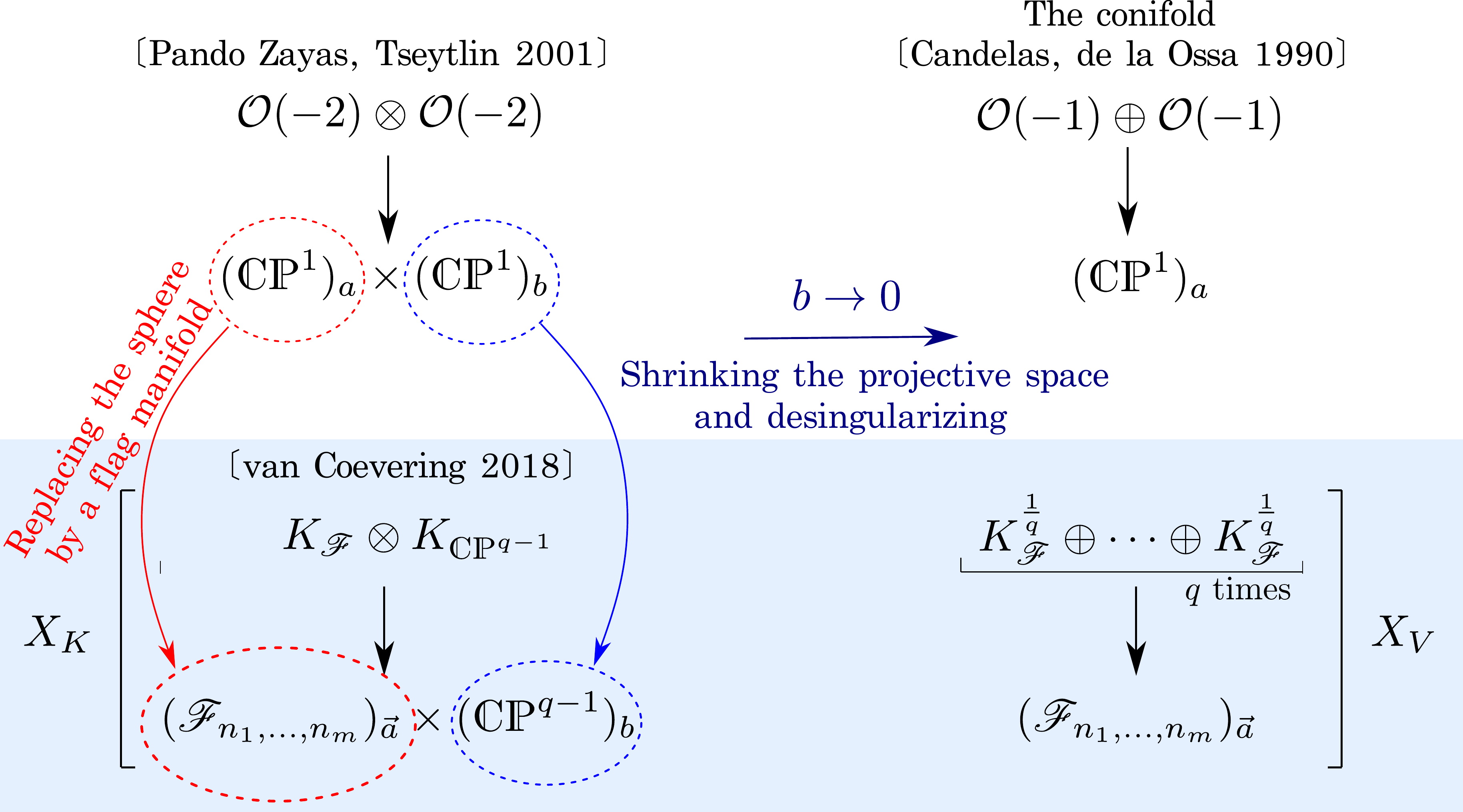}
    \caption{The lower (shaded) part of the picture depicts the idea of the present paper. We obtain both the metrics on $X_K$ and on $X_V$. The generalized vector $\vec{a}$ contains the K\"ahler moduli pertinent to the flag manifold, whereas $b$ is the size of the projective space $\CP^{q-1}$. `Desingularizing' means removing a $\CC^q/\mathbb{Z}_q$ orbifold singularity that arises in the limit $b\to 0$.}
    \label{scheme}
\end{figure}

The K\"ahler cone of the total space of the canonical bundle over the flag manifold is the same as that of the underlying flag manifold. The K\"ahler moduli of the flag manifold, in turn, can be easily characterized geometrically:
\begin{itemize}
	\item As parameters defining an adjoint orbit, in which case the K\"ahler form is the Kirillov-Kostant form on this orbit (see~\cite{Bykov:2012am} for a review).
	\item As Fayet-Iliopoulos parameters related to the gauged linear $\sigma$-model representations for flag manifolds~\cite{Nitta:2003dv,Donagi:2007hi}.
	\item The most general K\"ahler metric can be directly constructed using the so-called quasi-potentials~\cite{AzadKobayashiQureshi1997,AzadBiswas2003} that also featured in the physics literature in~\cite{BandoKugoYamawaki1988}. We will adopt this strategy throughout the paper.
\end{itemize}
 As a result, for the $s$-step flag manifold (i.e. for the flags of type~(\ref{flag1})) there are $s-1$ real moduli. Therefore Calabi's ansatz does not capture the full moduli space of Ricci-flat metrics on the total space. This was taken into account in~\cite{Coevering18}, where a full family of Ricci-flat metrics on the total space of $K_{\mathscr{F}}$ was constructed, using a generalization of Calabi's ansatz. In fact, this generalization is analogous to the one that arose in the work~\cite{CdO} on the conifold and was also considered in~\cite{Nitta:2003yk}.

Another interesting analogy may be observed, if one adds to the above the work~\cite{PZTmain}, where essentially the Ricci-flat K\"ahler metric on the total space of $K_{\CP^1\times \CP^1}$ was constructed. As required by the Calabi-Yau theorem, this metric has two K\"ahler moduli (since $H^2(\CP^1\times \CP^1, \mathbb{R})\simeq \mathbb{R}^2$), which geometrically correspond to the radii of the two spheres representing the `zero section'. As one of the spheres shrinks (i.e., as we approach the boundary of the K\"ahler cone in a particular way), one gets a metric on a $\CC^2/ \mathbb{Z}_2$ bundle over $\CP^1$. Such orbifold bundles were investigated in a more general context in~\cite{Martelli}. Removing the orbifold singularity at the zero section, one obtains the total space of the vector bundle $\mathcal{O}(-1)\oplus \mathcal{O}(-1)$ over $\CP^1$~\cite{CdO} (the so-called `conifold'). 

In the present paper we pursue a suitable generalization of this procedure to flag manifolds $\mathscr{F}_{n_1,\ldots, n_s}$. In this case, in place of $\CP^1\times \CP^1$, we start with the manifold $\mathscr{F}_{n_1,\ldots, n_s}\times \CP^{q-1}$, and construct a Ricci-flat metric on
\begin{equation}
X_K := \textrm{the total space of }\; K_{\mathscr{F}\times \CP^{q-1}}\,.
\end{equation}
In fact, alternatively one may view the manifold $\mathscr{F}_{n_1,\ldots, n_s}\times \CP^{q-1}$ as a flag manifold of a semi-simple group $SU(n)\times SU(q)$, which allows to identify these metrics with a special case of the metrics constructed in~\cite{Coevering18}. We then take the limit, when the volume of $\CP^{q-1}$ vanishes, remove a $\CC^q/\mathbb{Z}_q$-orbifold singularity and show that in the special case when the line bundle $K_{\mathscr{F}}^{1/q}$ is well-defined (i.e., when $q$ divides $\mathbf{c}_1(\mathscr{F}_{n_1,\ldots, n_s})$), the resulting manifold is
\begin{equation}
X_V:= \textrm{the total space of }\; \underbracket{K_{\mathscr{F}}^{1/q} \oplus ... \oplus K_{\mathscr{F}}^{1/q}}_{q \text{ times}}
\end{equation}
over $\mathscr{F}_{n_1,\ldots, n_s}$. The latter rank-$q$ vector bundle will be denoted $V$. The logic just described is summarized in Fig.~\ref{scheme}.

\vspace{0.3cm}
To formulate our result more precisely, we recall the expression for the first Chern class of the flag manifold:
\begin{equation}
\mathbf{c}_1(\mathscr{F}_{n_1,\ldots, n_s}) = -\sum\limits_{k=1}^{s-1} (n_k + n_{k+1}) \mathbf{c}_1(U_k),
\end{equation}
where $U_k$ are pullbacks of tautological bundles over $Gr(m_k,n)$ w.r.t. the forgetful projections $\pi_k: \mathscr{F}_{n_1,\ldots, n_s}\to Gr(m_k,n)$. We recall the derivation of this formula in section~\ref{canbunsect}.

\vspace{0.3cm}
\noindent We prove the following statement:

\vspace{0.3cm}
\noindent\textbf{Proposition.} \textit{There exists a complete Ricci-flat K\"ahler metric on $X_K$ in each K\"ahler class. If there exists a $q \in \mathbb{N}$ such that $q | (n_k + n_{k+1})$ (${k=1\ldots s-1}$), then there exists a complete Ricci-flat K\"ahler metric on $X_V$ in each K\"ahler class. In both cases the line element of the metric is of the form 
\begin{multline}
 \label{lineelem}
ds^2 =q (\alpha +  \mu) ds_{FS}^2 +
 \sum_{k=1}^{s-1} c_k (a_k + \mu) \pi_k^\ast\left(ds_k^2\right) +
 H_{\mu \mu} \frac{d \mu^2}{4} + {1\over H_{\mu \mu}}\left(d \phi + \mathrm{Im}(A)\right)^2,  
\end{multline}
where $ds_{FS}^2$ is the Fubini-Study metric on $\mathbb{CP}^{q-1}$, and $ds_k^2$ is the K\"{a}hler-Einstein metric, satisfying $R_{i\bar{j}}=n g_{i\bar{j}}$, on the Grassmannian manifolds $Gr(m_k,n)$, where $m_k =\sum_{i=1}^{k} n_i$. Besides, $c_k = n_k + n_{k+1}$, $A$ is the holomorphic connection of $K_{\mathscr{F}_{n_1,\ldots, n_s} \times \mathbb{CP}^{q-1}}$ and the constants $(a_k, \alpha)$ determine the K\"{a}hler class.  \\
The action of the complex structure $\mathscr{J}$ is given by $\mathscr{J} \left({1\over 2} H_{\mu\mu} d\mu\right)=d \phi + \mathrm{Im}(A)$.\\ 
\ \\
$H_{\mu \mu}$ is of the form
\bear
&& H_{\mu \mu} = \frac{F'(\mu)}{F(\mu)},\quad\quad\textrm{where}\\ \label{Ffunc}
&& F(\mu) = \int_{C}^{\mu} d\mu^{\prime} (\alpha + \mu^{\prime})^{q-1} \displaystyle \prod_{1 \leq i<j \leq s} \left( \sum_{k = i}^{j-1} c_k (a_k + \mu^{\prime}) \right)^{n_i n_j}.
\eear
If $\alpha + C >0$, $a_k + C >0$ for all $k$, and the angular variable $\phi$ takes values in $[0,2 \pi ]$ (\ref{lineelem})-(\ref{Ffunc}) describe a metric on $X_K$ for all K\"ahler classes.
\\ 
\ \\
If $a_k - \alpha >0$ for all $k$, $C = -\alpha$, and $\phi \in [0,2 \pi q]$ (\ref{lineelem})-(\ref{Ffunc}) describe a metric on $X_V$ for all K\"ahler classes.
\\
\ \\
The metrics so constructed are asymptotic to the Riemannian cone over the Sasaki-Einstein $U(1)$ bundles over $\mathscr{F}_{n_1,\ldots, n_s}\times \CP^{q-1}$. For $X_K$ and $X_V$ these are the unit vector bundles of $K_{\mathscr{F}_{n_1,\ldots, n_s} \times \mathbb{CP}^{q-1}}$ and its $q$-th root respectively.
}\\ \newline
\emph{Comment.} We note that the line element~(\ref{lineelem}) has the form of Pedersen-Poon~\cite{PP}.

\vspace{0.3cm}
The structure of the paper is as follows. In section~\ref{cp1cp1} we recall, how the Ricci-flat K\"ahler metric on $K_{\CP^1\times \CP^1}$ is constructed, using a generalization of Calabi's ansatz. 
In section~\ref{flagsec} we pass over to flag manifolds, starting in section~\ref{canbunsect} by explaining the expression for the first Chern class of a flag manifold and constructing the K\"ahler-Einstein metric in explicit form. Using the K\"ahler-Einstein metric, we construct in section~\ref{genCalabisect} a generalization of Calabi's ansatz, which allows obtaining the Ricci-flat metric in every K\"ahler class. This generalized ansatz leads to an ODE, which is then solved in section~\ref{genCalabisect}. In the same section the topology of the manifold (the behavior near the zero section, as well as at infinity) is also analyzed. The appendix is dedicated to the calculation of the determinant of the Hermitian metric, which is used in writing out the Ricci-flatness equation.

\section{Conifold and canonical bundle over $\mathbb{CP}^1 \times \mathbb{CP}^1$}\label{cp1cp1}

 The ansatz of Calabi may be succintly formulated as the requirement that the K\"ahler potential $\mathscr{K}$ on the total space assumes the form
\bea\label{CalabiAns}
\mathscr{K}=\mathscr{K}\left(|u|^2 \,e^{\widehat{K}}\right),
\eea
where $\widehat{K}$ is the K\"ahler potential of the K\"ahler-Einstein metric on the underlying manifold $\mathcal{M}_{KE}$ and $u$ is the coordinate in the fiber. Throughout the paper we will assume that the K\"ahler-Einstein metric $g_{KE}$ on the base is normalized so that
\bea \label{KahlerEinsteinEquation}
Ric=g_{KE}\,.
\eea

In this section as our principal example we will take $\mathcal{M}_{KE}=\CP^1\times \CP^1$. Note that this manifold has two K\"ahler moduli (the sizes of the two spheres), so this is the simplest instance of the situation described in the introduction, namely the Calabi-Yau theorem requires the existence of two parameters in the metric on $Y_K:= \;\textrm{the total space of}\; K_{\CP^1\times \CP^1}$. Calabi's ansatz~(\ref{CalabiAns}) fails to capture the full moduli space, since the K\"ahler-Einstein condition on $\CP^1\times \CP^1$ requires that the radii of the two $\CP^1$'s be equal.

Interestingly, Calabi's ansatz corresponds to a special point in the moduli space of metrics, namely the corresponding K\"ahler form $[\omega_{\textrm{Calabi}}]\in H^2_c(Y_K, \mathbb{R})$ lies in the \emph{compactly supported} cohomology. This is characterized by a faster decay to the asymptotic form at infinity. For a more detailed discussion of this see~\cite{Coevering, Goto, BykovPezzo}.

Denoting the inhomogeneous coordinates on the two spheres by $z$ and $w$, one introduces the following generalization of Calabi's ansatz (after a simple change of variables):
\bear\label{KahPZT1}
&&\mathscr{K}=a_1\,K_1+ a_2 \,K_2+ \mathscr{K}_0\left(\underbracket{|u|^2 e^{K_1+K_2}}_{:=x}\right),\\
&& K_1=2\log{(|z_1|^2+|z_2|^2)},\quad\quad K_2=2\log{(|w_1|+|w_2|^2)}
\eear
Clearly $K_1$ and $K_2$ are the K\"ahler potentials of the two spheres, and $K_1+K_2$ is the K\"ahler potential of the Einstein metric on $\CP^1\times \CP^1$. Setting $a_1=a_2=0$ would yield precisely the ansatz of Calabi.

Now, we are dealing with a toric variety, the $U(1)^3$ holomorphic isometries being given by the rotations
\bea
(z_1, z_2)\to  (e^{i\alpha_1 } z_1, e^{-i\alpha_1 } z_2), \quad (w_1, w_2)\to (e^{i\alpha_2 } w_1, e^{-i\alpha_2 }w_2), 
\quad u\to e^{i \beta} u.
\eea
In such cases it is useful to pass to the moment map variables. The moment map is defined as the derivative $\nabla_{\mathscr{J}v} \mathscr{K}$, where $\mathscr{J}$ is the complex structure, and $v$ is the vector field corresponding to the holomorphic isometry. In practical terms, this is tantamount to replacing $|z_1|^2+|z_2|^2\to e^{s_1}+e^{-s_1}$, $|w_1|^2+|w_2|^2\to e^{s_2}+e^{-s_2}$, $|u|^2\to e^t$ in the K\"ahler potential, and differentiating it w.r.t. $s_i$ and $t$:
\bea
\mu=\frac{\dd \mathscr{K}}{\dd t}=x \mathscr{K}_0'(x),\quad \nu_i=\frac{\dd \mathscr{K}}{\dd s_i}=2\,\mathrm{th}(s_i)\,(a_i+\mu) 
\eea
One also computes the so-called symplectic potential $\mathscr{H}$, which is the Legendre transform of the K\"ahler potential $\mathscr{K}$ w.r.t. the variables $t, s, p$:
\bea
\mathscr{H}(\mu, \nu, \mu_3)= t\mu+s_1\nu_1+s_2\nu_2-\mathscr{K}
\eea
The nice feature of this potential is that the domain, on which it is defined, is precisely the moment polytope of the manifold. Clearly, since in our case the manifold is non-compact, the polytope is also unbounded.

Up to inessential linear terms in $\mu$, the dual potential reads:
\bear\label{fullH}
\mathscr{H}=H(\mu)+\sum\limits_{i=1}^2\,\ell_i^+ \log{\ell^+_i}+\sum\limits_{i=1}^2\,\ell^-_i \log{\ell^-_i} -2\sum\limits_{i=1}^2\,(\mu+a_i)\log{(\mu+a_i)},\\ \nonumber
\textrm{where}\quad\quad H(\mu)=\mu \log{x}-\mathscr{K}_0\,,\quad\quad \ell_i^\pm=\mu+a_i\pm {1\over 2}\nu_i\,.
\eear
Note that the $l \log(l)$ structures appearing in the symplectic potential are typical for K\"{a}hler toric geometry \cite{guillemin1994}. 
Now, the function $H(\mu)$ is determined from the Ricci-flatness equation, which in this case reads
\bea
3 (\mu+a_1)(\mu+a_2)=\frac{d}{d\mu}\left(e^{H_\mu}\right)\,.
\eea
The solution is $H=\sum\limits_{i=0}^2\,(\mu-\mu_i) (\log{(\mu-\mu_i)}-1)$, where $\mu_i$ are the three roots of the polynomial $P(\mu)=\int\limits_{\mu_0}^\mu\,d\mu'\ \,3 (\mu'+a_1)(\mu'+a_2)$.

Next we write down the explicit expression for the line element derived from the K\"ahler potential~(\ref{KahPZT1}), using the dual variable $\mu$:
\bear\nonumber
&&\!\!\!\!\!\!\!\!\!\!ds^2=(\mu+a_1)\,(ds^2)_{\CP^1_z}+(\mu+a_2)\,\,(ds^2)_{\CP^1_w}+{1\over 4} H_{\mu\mu} d\mu^2+{1\over H_{\mu\mu}}\,(d\phi +\mathrm{Im}(A))^2,\\
&&(ds^2)_{\CP^1_z}=\frac{2\,dz d\bar{z}}{(1+|z|^2)^2},\quad  (ds^2)_{\CP^1_w}=\frac{2\,dw d\bar{w}}{(1+|w|^2)^2}\,,\\ \nonumber &&A=2\left(\frac{ \bar{z} dz}{1+|z|^2}+\frac{\bar{w} dw}{1+|w|^2}\right)\,.
\eear
Here $A$ is the holomorphic connection on the canonical bundle $K_{\CP^1\times \CP^1}$. Non-negativity of the metric implies $\mu+a_1\geq 0, \mu+a_2\geq 0, H_{\mu\mu}=\frac{P'(\mu)}{P(\mu)}\geq 0$. The latter requirement is equivalent to $P(\mu)\geq 0$. Note that $P(\mu_0)=0$ and $P(\mu)>0$ for $\mu>\mu_0$, due to the first two inequalities. Completeness requires that   the range of $\mu$ is $\mu\in[\mu_0, \infty)$. In fact, there are two distinct possibilities:
\begin{itemize}
\item $\mu_0>\mathrm{max}(-a_1, -a_2)$. This was considered by Pando Zayas and Tseytlin~\cite{PZTmain}.
\item $\mu_0=\mathrm{max}(-a_1, -a_2)$. This was considered in the early work of Candelas and de la Ossa~\cite{CdO}.
\end{itemize}
In the first case, one has $H_{\mu\mu}=\frac{1}{\mu-\mu_0}+\ldots$ as $\mu\to\mu_0$. The change of variables $r=(\mu-\mu_0)^{1/2}$ brings the metric to the asymptotic form
\bea\label{asympmetr1}
(ds^2)_{\mu\to \mu_0}=(\mu_0+a_1)\,(ds^2)_{\CP^1_z}+(\mu_0+a_2)\,\,(ds^2)_{\CP^1_w}+dr^2+r^2\,(d\phi+\mathrm{Im}(A))^2\,.
\eea
The absence of singularity at $r=0$ requires that the angle $\phi$ has the range $\phi\in[0, 2\pi]$. Then (\ref{asympmetr1}) shows that one has the canonical bundle, with connection $\mathrm{Im}(A)$, over $\CP^1\times \CP^1$, the two spheres having radii squared $\mu_0+a_1$ and $\mu_0+a_2$. Varying $\mu_0$ leads to changing the K\"ahler class of the metric, and all allowed classes (corresponding to the non-vanishing sizes of the $\CP^1$'s) can be achieved in this way.

\begin{figure}[h]
    \centering
    \includegraphics[width=\textwidth]{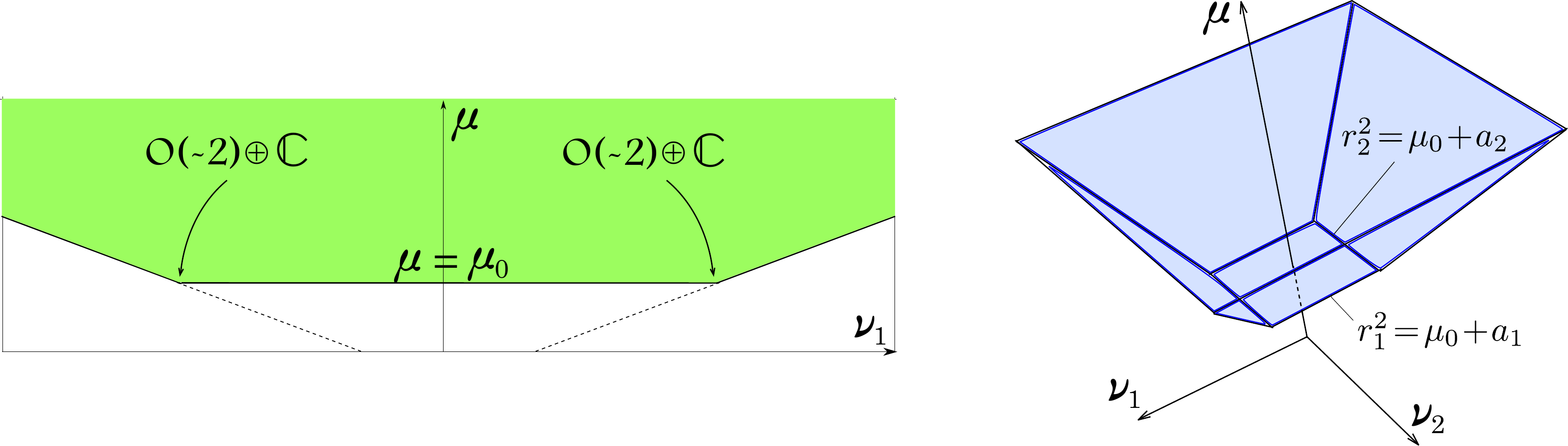}
    \caption{The moment polytope for the cone over $\CP^1 \times \CP^1$ (right) and its section in the $(\mu, \nu_1)$ plane (left). The radii of the two spheres representing the zero section are denoted $r_1, r_2$.}
    \label{PZTpyr}
\end{figure}
\begin{figure}[h]
    \centering
    \includegraphics[width=\textwidth]{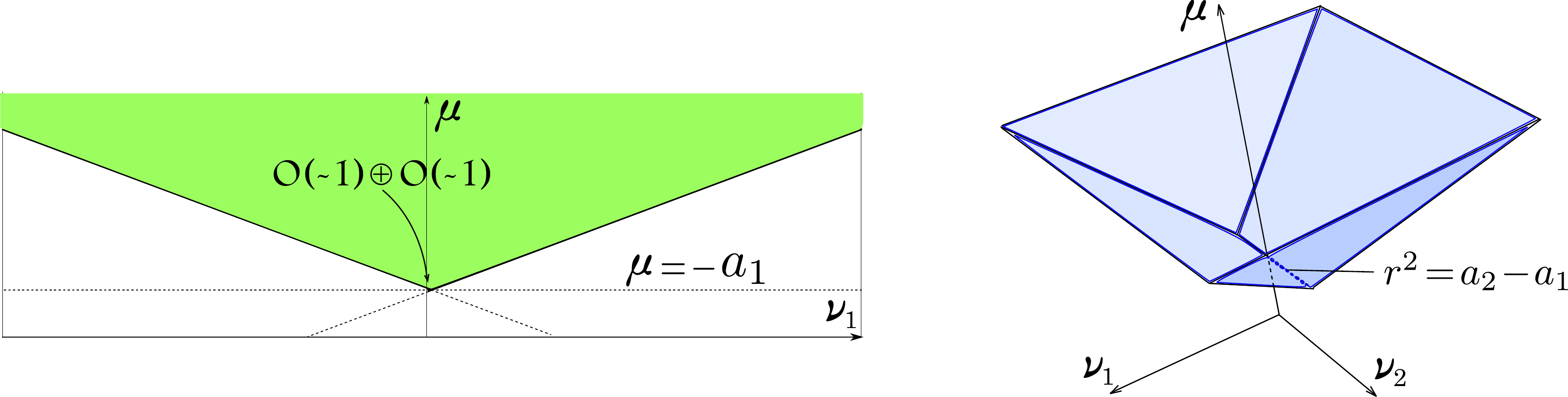}
    \caption{The moment polytope for the total space of $\mathcal{O}(-1)\oplus \mathcal{O}(-1)$ vector bundle over $\CP^1$ (right) and its section in the $(\mu, \nu_1)$ plane (left). The radius of the sphere representing the zero section is denoted $r$.}
    \label{CdOpyr}
\end{figure}

In the second case, let us assume that $\mathrm{max}(-a_1, -a_2)=-a_1$. $P(\mu)$ now has a second-order zero at $\mu=\mu_0$, so that $H_{\mu\mu}=\frac{2}{\mu-\mu_0}+\ldots$ Introducing the variable $r=(2(\mu-\mu_0))^{1/2}$, in the vicinity of $\mu=\mu_0$ we can bring the metric to the form
\bea\label{zerosectOmin1}
(ds^2)_{\mu\to \mu_0}=(a_2-a_1)\,(ds^2)_{\CP^1_z}+dr^2+r^2\,\left(\frac{dw d\bar{w}}{(1+|w|^2)^2}+\left(d\tilde{\phi}+{1\over 2}\mathrm{Im}(A)\right)^2\right)\,.
\eea
Here $\tilde{\phi}={\phi\over 2}$. If one sets $z=\mathrm{const.}$, the expression in round brackets is precisely the round metric on $S^3$, provided one takes the periodicity of the angle $\tilde{\phi}$ to be $\tilde{\phi}\in[0, 2\pi]$. In this case the conical part $dr^2+r^2(\ldots)$ of the metric above describes the space $\mathbb{R}^4\simeq \CC^2$. (Keeping instead the periodicity $\phi\in[0, 2\pi]$ would lead to a $\CC^2/\mathbb{Z}_2$ orbifold singularity at the zero section, $r=0$.) Taking into account the stereographic $z$-variable of the remaining sphere $\CP^1$, one can show that the metric~(\ref{zerosectOmin1}) describes the vicinity of the zero section in the $\mathcal{O}(-1)\oplus \mathcal{O}(-1)$ bundle over $\CP^1$ (the so-called~`conifold')~\cite{CdO}.

Since the varieties we have considered are toric, it is instructive to construct the respective toric polytopes. The polytopes corresponding to the Pando Zayas-Tseytlin solution and the Candelas-de la Ossa solution are schematically presented in Figs.~\ref{PZTpyr} and \ref{CdOpyr}, respectively. The sections of the moment polytope in the $(\mu, \nu)$ plane are shown to the left of the full three-dimensional polytope.

\section{Flag manifolds}\label{flagsec}

In this section, apart from the unitary representation~(\ref{flag2}) for the flag manifold~(\ref{flag1}), it will be useful to recall the complex parametrization $\mathscr{F}_{n_1,\ldots, n_s} = Gl(n;\CC)/P_{n_1,\ldots, n_s}$ (with $P_{n_1,\ldots, n_s}$ a parabolic subgroup, stabilizing a given flag), which expresses the flag manifold as a homogeneous space of the complex linear group $Gl(n;\CC)$. 

We will start by recalling the explicit form of the K\"ahler-Einstein metric on the flag manifold, which is an essential ingredient in Calabi's ansatz. We then use a suitable generalization of this ansatz to construct the metric on $X_K$ -- the total space of the canonical bundle over $\mathscr{F}_{n_1,\ldots, n_s} \times \mathbb{CP}^{q-1}$, and on $X_V$ -- the total space of the vector bundle $V := \underbracket{K_\mathscr{F}^{1/q} \oplus ... \oplus K_\mathscr{F}^{1/q}}_{q \text{ times}}$ (the latter in the case when the $q$-th root of $K_\mathscr{F}$ makes sense).

\subsection{The canonical bundle}\label{canbunsect}

On $\mathscr{F}_{n_1,\ldots, n_s}$ we can consider the vector bundles $\xi_j$ and $U_j$ ($j = 1,...,s$) where the fiber of $\xi_j$ over the point
\begin{equation}
0 \subset L_{1} \subset ... \subset L_{s-1} \subset L_s=\mathbb{C}^{n}
\end{equation}
is given by $L_{j}/L_{j-1}$, and the fiber of $U_j$ is $L_j$ ($U_j$ are the tautological bundles). Here, as before, $L_{k} \cong \CC^{m_k}$. As is well-known \cite{Lam75}
\begin{equation}
T\mathscr{F}_{n_1,\ldots, n_s} = \bigoplus_{1 \leq i < j \leq s} \xi_i^{\ast} \otimes \xi_j.
\end{equation}
We will be interested in the explicit expression for the first Chern class of the flag manifold:
\begin{eqnarray}
\mathbf{c}_1(\mathscr{F}_{n_1,\ldots, n_s}) = \sum_{i<j} (- n _j \mathbf{c}_1(\xi_i) + n_i \mathbf{c}_1(\xi_j)).
\end{eqnarray}
Rewriting in terms of $U_j$ gives
\begin{equation}\label{firstChernclass}
\mathbf{c}_1(\mathscr{F}_{n_1,\ldots, n_s}) = - \sum_{k=1}^{s-1} (n_k + n_{k+1}) \mathbf{c}_1(U_k).
\end{equation}
Now, suppose there is a positive integer $q$ that divides $(n_k + n_{k+1})$ for all $k$. Then, since $\mathrm{Pic}(\mathscr{F})\simeq H^{2}(\mathscr{F}, \mathbb{Z})$~\cite[Prop. 2.1.2]{AGV}\cite{brion2003}, there is a line bundle $K_\mathscr{F}^{1/q}$, where $K_\mathscr{F}$ is the canonical bundle of $\mathscr{F}_{n_1,\ldots, n_s}$.

Furthermore since $\mathscr{F}_{n_1,\ldots, n_s}$ is K\"{a}hler we have that the Ricci-form represents $\mathbf{c}_1(\mathscr{F}_{n_1,\ldots, n_s})$. Let $(u_1,...,u_n) \in U(n)$, where $u_i$ are column vectors. Then 
\begin{equation}\label{c1Uk}
\mathbf{c}_1(U_k) = \left[ \frac{1}{2 \pi i} \sum_{i=1}^{m_k} \sum_{j=m_k+1}^{n} J_{ij}\wedge J_{ji} \right],
\end{equation}
with $J_{ij} = \sum_m \bar{u}_{jm} du_{im}$. Therefore the line element of the K\"{a}hler-Einstein metric on $\mathscr{F}_{n_1,\ldots, n_s}$, satisfying \eqref{KahlerEinsteinEquation}, must take the form 
\begin{equation} \label{generalKEMetricOnF}
ds^2 = \sum_{k=1}^{s-1} (n_k + n_{k+1}) \displaystyle \sum_{i=1}^{m_k} \sum_{j=m_k+1}^n |J_{ij}|^2.
\end{equation}
Note that K\"ahler-Einstein metric on flag manifolds were first discussed in \cite{AP}. It is useful for the following discussion to write out explicitly the K\"ahler potential corresponding to this metric. To this end consider the matrix
\begin{equation}\label{Wmat}
W=(w_1,..,w_n) \in Gl(n;\mathbb{C}),
\end{equation}
where each $w_i$ is a column vector. We also define an $n \times m_k$-matrix $Z_k \in \mathrm{Hom}(\CC^{m_k}, \CC^n)$ of rank $m_k$
\begin{equation}
Z_k = (w_1,...,w_{m_k}), \ \ \ \text{where } \ m_k = \sum_{l=1}^{k} n_l\,,
\end{equation}
and introduce the function
\begin{equation}
t_k =   \det \left( Z_k^\dagger Z_k  \right)\,.
\end{equation}
One can check that $\log(t_k)$ is the K\"ahler potential for the $\pi$-normalized canonical metric\footnote[1]{The same normalization as that of the Fubini-Study metric on $\mathbb{CP}^{n-1}$, i.e. the volume of a holomorphic 2-sphere generating $H_2(Gr(m_k, n),\mathbb{Z})$ is $\pi$.} on the Grassmannian $Gr(m_k, n)$. The potential of an arbitrary K\"ahler metric on the flag manifold~\cite{AzadKobayashiQureshi1997,AzadBiswas2003} may then be written as~
\bea \label{KahlerPotFlagMfds}
\mathscr{K}_{\mathscr{F}}=\sum\limits_{k=1}^{s-1}\,\gamma_k\,\log(t_k)\,,\quad\quad \gamma_k>0\,.
\eea
In particular the K\"ahler-Einstein metric \eqref{generalKEMetricOnF} arises when $\gamma_{k}=n_k+n_{k+1}:=c_k$.

\subsection{Generalized Calabi Ansatz for all K\"{a}hler classes}\label{genCalabisect}

From now on we assume there is a positive integer $q$ that divides $c_k:=n_k + n_{k+1}$ for all $k=1,\ldots, s-1$. Also we define the vector $\vec{z} = (z_1,...,z_{q-1}) \in \CC^{q-1}$.\\
As a candidate for a K\"{a}hler potential on~$X_K$, we define 
\begin{eqnarray}\label{KahlerPotential}
\mathscr{K} = \alpha q \log (1 + |\vec{z}|^2) + \sum_{k=1}^{s-1} a_k c_k \log t_k + \mathscr{K}_0(\underbracket{|u|^2 (1 + |\vec{z}|^2)^q \Pi_k t_k^{c_k}}_{=: e^t}).
\end{eqnarray}
From the discussion in the previous section it follows that $\widehat{K}=q \log(1 + |\vec{z}|^2)+\sum\limits_{k=1}^{s-1}\,c_k\,\log(t_k)$ is the K\"ahler potential of the K\"ahler-Einstein metric on $\mathscr{F}_{n_1,\ldots, n_s} \times \mathbb{CP}^{q-1}$, and therefore the last term in~(\ref{KahlerPotential}) is the expression familiar from Calabi's ansatz~(\ref{CalabiAns}).

In (\ref{KahlerPotential}) $(\alpha,a_k)$ are the parameters (K\"ahler moduli), akin to $a_1, a_2$ from (\ref{KahPZT1}).  Their range will be determined later. If we work on $X_K$, the components $z_i$ of  $\vec{z}$ are local coordinates on $\mathbb{CP}^{q-1}$ and $u$ is a holomorphic coordinate on the fiber. If we work on $X_V$, $(u,z_i)$ are local coordinates on the fiber.

Notice that $\mathscr{K}_0$ depends only on a single variable $t$. We will write $\mathscr{K}_0^{\prime}$ and $\mathscr{K}_0^{\prime \prime}$ for its first and second derivatives w.r.t $t$. In fact, instead of dealing with the function $\mathscr{K}_0$, just like in section~(\ref{cp1cp1}) it is convenient to perform a Legendre transform
\begin{equation}
H = \mu t - \mathscr{K}_0,
\end{equation}
whence
\begin{equation}
\mathscr{K}_0^{\prime} = \mu, \ \ \ \mathscr{K}_0^{\prime \prime} 
= \frac{1}{ H_{\mu \mu}}.
\end{equation}
The meaning of $\mu$ is that it is the moment map for the $U(1)$-action $u\to e^{i\alpha} u$.

The line element then is given by  
\begin{eqnarray}\label{LineElement}
ds^2 = q(\alpha + \mu) ds_{FS}^2  + \sum_{k = 1}^{s-1} c_k (a_k + \mu) \pi_k^{\ast} (ds_k^2) +
 \frac{1}{|u|^2 H_{\mu \mu}}  \ |du + u A|^2,
\end{eqnarray}
where $ds_{FS}^2$ has the form of a Fubini-Study metric on $\mathbb{CP}^{q-1}$ and
\begin{eqnarray}\label{GeneralConnection}
&\pi_k^{\ast} (ds_k^2) &= (\partial_i \bar{\partial}_j \log t_k) dy^i d \bar{y}^j \\
& A &= \sum_{i,k} c_k \partial_i \log t_k dy^i + q \partial_i \log(1 + |\vec{z}|^2) dz^i.
\end{eqnarray}
Note that $y^i$ are the complex coordinates on the flag manifold, and are abbreviations for $w_{lk}$ (the components of the vectors $w_l$). 

The expression~(\ref{LineElement}) may be brought to the form~(\ref{lineelem}) used in the statement of the Proposition, if we introduce the angular variable $\phi=\mathrm{arg}(u)$. Then, again using~(\ref{tdef}), we obtain
\begin{eqnarray}
\log u = \frac{t}{2} + i \phi - \frac{q}{2} \log (1 + |\vec{z}|^2) - \frac{c_k }{2} \sum_k \log t_k. 
\end{eqnarray}
Since $t = H_{\mu}$, we arrive at
\begin{eqnarray}
\frac{du}{u} + A  = d \log u + A = \frac{H_{\mu \mu} d \mu}{2} + i D \phi,
\end{eqnarray}
with $D \phi = d \phi + \text{Im}(A)$ as covariant derivative. Since the l.h.s. of the above equality is a one-form of type $(1,0)$, on which the complex structure $\mathscr{J}$ acts by multiplication by $-i$, we get
\bea
\mathscr{J}\left(\frac{H_{\mu \mu} d \mu}{2}\right)=D \phi\,.
\eea
The $u$-dependent part of the metric~(\ref{LineElement}) is then
\begin{eqnarray}\label{polarpart}
\frac{1}{ H_{\mu \mu}} \big|\frac{du}{u} + A\big|^2  = H_{\mu \mu } \frac{d \mu^2}{4} + \frac{(D \phi)^2}{ H_{\mu \mu}}.
\end{eqnarray}

We denote by $g$ the Hermitian (K\"{a}hler) metric corresponding to \eqref{LineElement}. The Ricci-tensor is given by $
R_{i \bar{j}} = -\partial_{i} \bar{\partial}_{j} \log \det g\,,
$
where $\partial_i$ are derivatives w.r.t. the holomorphic coordinates. Therefore Ricci-flatness is satisfied if
\bea\label{RicciflatDeterminant}
\det g \sim \kappa \bar{\kappa}
\eea
for a holomorphic function $\kappa$.
It will be shown in App.~\ref{Appendix} that 
\begin{equation}\label{DeriveODE}
\det g =  \big|\kappa(\{w_{mi}\})\big|^2\,\frac{1}{|u|^2 H_{\mu  \mu} } \frac{q^{q-1}(\alpha + \mu)^{q-1}}{(1 + |\vec{z}|^2)^q} \frac{1}{ \prod_k t_k^{c_k}}  f(\mu),
\end{equation}
where $\kappa$ is a holomorphic function, and
\begin{equation}\label{ffunc}
f(\mu) = \displaystyle \prod_{1 \leq i<j \leq s} \left(\sum_{k= i}^{j-1} c_k (a_k + \mu) \right)^{n_i n_j}.
\end{equation} 
\ \\
Using the definition
\bea\label{tdef}
e^t=|u|^2 (1 + |\vec{z}|^2)^q \displaystyle \prod_{k} t_k^{c_k}
\eea
from \eqref{KahlerPotential},
as well as $H_{\mu} = t$, we may satisfy \eqref{RicciflatDeterminant} by requiring
\begin{equation}\label{ODE}
(\alpha + \mu)^{q-1} f(\mu) =  H_{\mu \mu} \exp( H_{\mu} ) = \partial_{\mu} \exp( H_{\mu}).
\end{equation}

\subsection{Solution of the Ricci-flatness equation}\label{ODEsolsec}

The ODE~(\ref{ODE}) is solved by
\begin{equation}\label{Hmu}
H_{\mu} =  \log \left( \int_{C}^{\mu} d\mu^{\prime} (\alpha + \mu^{\prime})^{q-1} f(\mu^{\prime}) \right).
\end{equation}
Therefore 
\begin{equation}\label{Hmumu}
H_{\mu \mu} = \frac{(\alpha + \mu)^{q-1} f(\mu)}{\int_{C}^{\mu} d\mu^{\prime} (\alpha + \mu^{\prime})^{q-1} f(\mu^{\prime})}.
\end{equation}

\subsubsection{Behavior at $\infty$}
Before analysing the effect of different choices for the integration constant $C$, we observe that 
\begin{equation}
H_{\mu \mu} \rightarrow \frac{N}{\mu}\quad\quad \text{ as} \quad\quad \mu \rightarrow \infty.
\end{equation}
Here $N = q + \mathrm{dim}_{\CC}(M)$. Using (\ref{LineElement}) and (\ref{polarpart}) and making the substitution $\mu = {1\over N}r^2$, we find that the line element behaves at infinity as 
\begin{eqnarray}
ds^2|_{\mu \rightarrow \infty}  = dr^2 + r^2\left({1\over N} ds_{KE}^2+\frac{1}{N^2} (D \phi)^2 \right)\,,
\end{eqnarray}
where $ds_{KE}^2=q ds_{FS}^2  + \sum_{k = 1}^{s-1} c_k  \pi_k^{\ast} (ds_k^2)$ is the K\"ahler-Einstein metric on $\mathscr{F}_{n_1,\ldots, n_s} \times \mathbb{CP}^{q-1}$ with proportionality factor $1$.
Thus at $\mu \rightarrow \infty$ we obtain a cone over a Sasakian manifold, which is a $U(1)$-bundle over $\mathscr{F}_{n_1,\ldots, n_s} \times \mathbb{CP}^{q-1}$. As will be demonstrated in the remainder of this section the angle $\phi$ ranges from $[0, 2 \pi]$ on $X_K$ and from $[0,2 \pi q]$ on $X_V$. Therefore the $U(1)$-bundles are the unit -vector bundles of $K_{\mathscr{F}_{n_1,\ldots, n_s} \times \mathbb{CP}^{q-1}}$ and $K_{\mathscr{F}_{n_1,\ldots, n_s} \times \mathbb{CP}^{q-1}}^{1/q} $ respectively.

\subsubsection{The metric on $X_K$}
We now analyze the dependence on the choice of $C$.
If $C$ is larger than the largest root of $(\alpha + \mu)^{q-1} f(\mu)$, then $H_{\mu \mu }$ is strictly positive. The only possible issue would be a singularity at the zero section, but since
\begin{eqnarray}\label{HmumuCanonicalbundle}
H_{\mu \mu} = \frac{1}{\mu -C} + O(1), \quad\quad \text{as } \quad\quad \mu \rightarrow C,
\end{eqnarray}
the substitution $r^2 = \mu -C$ implies
\begin{equation}
\frac{1}{ H_{\mu \mu}} |\frac{du}{u} + A|^2  = dr^2 + r^2 (D \phi)^2 + ...,
\end{equation}
that is to say the line element corresponds to a smooth metric. The complex variable $r e^{i \phi}$ parametrizes $\CC$ -- the fiber of a line bundle, and the connection $A$ allows to identify this bundle with $K_{\mathscr{F}\times \CP^{q-1}}$. Therefore the underlying manifold is $X_K$.
The condition
\begin{equation}
a_k + C > 0, \ \ \ \ \alpha + C >0,
\end{equation}
makes the metric positive-definite. Upon restricting to the zero section $\mu = C$, \eqref{LineElement} reduces to 
\begin{equation}
ds^2 = \gamma_0 \,ds_{FS}^2 + \sum \gamma_k \,\pi_k^{\ast} (ds_k^2),
\end{equation}
with $\gamma_0 = q(\alpha + C), \gamma_k = c_k (a_k + C)$. The definition~\eqref{KahlerPotFlagMfds} of the K\"ahler cone of the flag manifold, combined with $\gamma_0>0$ for $\CP^{q-1}$, implies that all K\"ahler classes are obtained.

\subsubsection{The metric on $X_V$}
We will now show that for $C = -\alpha$ the line element \eqref{LineElement} describes a metric on $X_V$. Notice that positive-definiteness of the metric \eqref{LineElement4} requires (since $c_k>0$ for all $k$)
\begin{equation}\label{EuclideanConditionOnV}
a_k - \alpha >0.
\end{equation}
It then follows easily from~(\ref{ffunc}),~(\ref{Hmumu}) that the above condition also makes $H_{\mu \mu}$ (and hence the metric~(\ref{lineelem})) positive for $\mu > -\alpha$.

One can visualize the limit $\mu \rightarrow -\alpha$ as sending the volume of $\mathbb{CP}^{q-1}$ to zero and embedding $\mathbb{CP}^{q-1}$ into the fiber $\CC^q$. First we notice that
\bea\nonumber
\int_{-\alpha}^{\mu} d\mu^{\prime} (\alpha + \mu^{\prime})^{q-1} f(\mu^{\prime}) = \delta (\alpha+\mu)^q\left(1+O(\alpha+\mu)\right),\quad\quad \delta=\frac{f(-\alpha)}{q} > 0\,,
\eea
Since $t=H_\mu$, \eqref{Hmu} yields
\bea\label{tasympt}
e^t= \delta (\alpha+\mu)^q\left(1+O(\alpha+\mu)\right)\,.
\eea
Therefore to leading order in $\alpha+\mu$
\bear
&&\mathscr{K}_0=\mu H_\mu-H=\\ \nonumber
&&=\mu \log\left(\delta (\alpha+\mu)^q\right)-\log\left(\delta \right)\,(\mu+\alpha)-q (\mu+\alpha)(\log(\mu+\alpha)-1)+\mathrm{const.}=\\ \nonumber
&&=-q\alpha\log(\mu+\alpha)+q(\mu+\alpha)+\mathrm{const.}
\eear
Inserting this in the full K\"ahler potential (\ref{KahlerPotential}) and using~(\ref{tasympt}), we get (up to an additive constant that we forget from now on)
\bea
\mathscr{K}=\sum_{k=1}^{s-1} (a_k-\alpha) c_k \log t_k + B\,\left(\sum\limits_{m=0}^{q-1}\,|x_m|^2\right) \prod_k t_k^{c_k\over q}+\ldots,\quad\quad B=\frac{q}{\delta^{1\over q}}\,.
\eea
Here we have introduced the coordinates $(x_0,\cdots,x_{q-1})$ on the $\CC^{q}$ fiber. They are related to $(u, \vec{z})$ via $x_0=u^{1\over q}, x_i=u^{1\over q} z_i \;\;(i>0)$. This procedure changes the periodicity of $\mathrm{arg}(u)$ w.r.t the metric on $X_K$ in complete analogy to the Pando Zayas-Tseytlin/Candelas-de la Ossa metrics discussed in section \ref{cp1cp1} (keeping the original periodicity would result in a $\CC^q/\mathbb{Z}_q$-singularity). 

As we will now see, the above formula provides a K\"ahler potential in the vicinity of the zero section $\mathscr{F}_{n_1, \ldots, n_m}\subset X_V$. The zero section is given by the equations $x_m=0$, $m=0, \ldots q-1$.

Introducing the holomorphic connection $\widehat{A}=\sum\limits_k \frac{c_k}{q}\,\dd\log(t_k)$, we may write out explicitly the line element corresponding to the above potential, in the limit $x_m\to 0$:
\begin{eqnarray}\label{LineElement3}
ds^2 =  \sum_{k = 1}^{s-1} c_k (a_k -\alpha) ds_k^2 +
 \prod_k t_k^{c_k\over q}\sum\limits_{m=0}^{q-1}\left|dx_m+\widehat{A} x_m\right|^2+\ldots
 \end{eqnarray}
 The second term gives a metric in the fiber of the rank-$q$ vector bundle $V=\underbracket{K_{\mathscr{F}}^{1/q} \oplus ... \oplus K_{\mathscr{F}}^{1/q}}_{q \text{ times}}$. 
 
At the expense of introducing a non-holomorphic complex coordinate $\tau_m=\left(\prod_k t_k^{c_k\over q}\right)^{1/ 2} x_m=\rho_m e^{i\phi_m}$, we may rewrite~(\ref{LineElement3}) more compactly:
 \begin{eqnarray}\label{LineElement4}
&&ds^2 =  \sum_{k = 1}^{s-1} c_k (a_k -\alpha) ds_k^2 +
\sum\limits_{m=0}^{q-1} \left|d\tau_m+i\,\mathrm{Im}(\widehat{A}) \tau_m\right|^2+\ldots=\\ \nonumber
&& = \sum_{k = 1}^{s-1} c_k (a_k -\alpha) ds_k^2 +
\sum\limits_{m=0}^{q-1} \left(d\rho_m^2+\rho_m^2 (d\phi_m+\mathrm{Im}(\widehat{A}))^2\right)+\ldots
\end{eqnarray}
The absence of a conical singularity at $\rho_m=0$ implies that $\phi_m$ is periodic in the segment $[0, 2\pi]$.

\vspace{0.3cm}
\textbf{Acknowledgments.} We would like to thank D.~L\"{u}st and A.~A.~Slavnov for support and D.~Ageev, O.~Biquard, P.~Gauduchon, S.~Gorchinskiy, M.~Nitta, K.~Shramov, P.~Zinn-Justin for useful discussions. The research of I.A. was supported by the IMPRS program of the MPP Munich.

\vspace{1cm}
\appendix

\section[Calculation of the determinant of the metric]{\large Calculation of the determinant of the metric}\label{Appendix}

The goal of this section is to compute the determinant of the Hermitian metric corresponding to the line element~(\ref{LineElement}).

First we notice that it is easy to factor out the pieces corresponding to the $u$-dependence of the metric, as well as to the $\CP^{q-1}$ directions. As a result we get
\begin{equation}
\det g = \frac{1}{|u|^2 H_{\mu \mu}} \frac{(q (\alpha +  \mu))^{q-1}}{(1 + |\vec{z}|^2)^q} \det\left(g_{\mathrm{red}}^{(\mu)}\right)\,,
\end{equation}
where $g_{\mathrm{red}}^{(\mu)}$ is the reduced Hermitian metric, with the line element  
\begin{equation}\label{LineElementOnM}
ds_\mathscr{F}^2 = \sum_{k=1}^{s-1} c_k (a_k + \mu) \pi_k^{\ast}(ds_k^2)\,.
\end{equation}
This is a family of metrics on the flag manifold, depending on $\mu$ as a parameter. We will calculate the determinant by studying the set of values of $\mu$, for which $g_{\mathrm{red}}^{(\mu)}$ degenerates. Since the flag manifold is a homogeneous space, and the metric is a $SU(n)$-invariant tensor,  $\mathrm{rank}(g_{\mathrm{red}}^{(\mu)})$ is the same at every point. To calculate it, we consider the open set in $\mathscr{F}$, where the matrix $W$ introduced in~(\ref{Wmat}), which represents the quotient $Gl(n, \CC)/P$, may be brought to lower-block-triangular form. Evaluating $\pi_k^\ast(ds^2_k)$ at the point $W=1$, we obtain:
\begin{equation}
\pi_k^\ast(ds^2_k) \big|_{W=1} = \sum_{j= m_k + 1}^{n} \sum_{i = 1}^{m_k} |d w_{ji}|^2.
\end{equation}
Thus 
\begin{eqnarray}
\nonumber
& ds_\mathscr{F}^2 \big|_{W=1}  & = \sum_{k = 1}^{s-1}	c_k (a_k + \mu) \sum_{j=m_k+1}^{n} \sum_{i= 1}^{m_k} |dw_{ji}|^2 \\
&& \nonumber =  \sum_{i= 1}^{s-1} \sum_{j = i+1}^{s} \left( \sum_{k= i}^{j-1} c_k (a_k + \mu) \right) \sum_{a= m_{j-1} + 1}^{m_{j}} \sum_{b= m_{i-1} + 1}^{m_{i}} |dw_{ab}|^2.
\end{eqnarray}
The metric at $W=1$ is therefore diagonal, with eigenvalues being equal to $  \sum_{k= i}^{j-1} c_k (a_k + \mu)$, each of multiplicty $n_i n_j$. At an arbitrary point $W\in \mathscr{F}$, the Hermitian metric $g_{\mathrm{red}}^{(\mu)}$ is represented by a matrix of size $\mathrm{dim}_\CC \mathscr{F}\times \mathrm{dim}_\CC \mathscr{F}$, whose entries are linear in $\mu$. Since $\sum\limits_{i<j}\,n_i n_j=\mathrm{dim}_\CC \mathscr{F}$, we get
\begin{equation}
\det(g_{\mathrm{red}}^{(\mu)}) =\xi\left(\{w_{mi}, \bar{w}_{mi}\}\right) f(\mu),
\end{equation}
where
\begin{equation}
f(\mu)=\det g_{\mathrm{red}}^{(\mu)}\big|_{W=1}  = \displaystyle \prod_{1 \leq i<j \leq s} \left( \sum_{k= i}^{j-1} (a_k + c_k \mu) \right)^{n_i n_j}\,
\end{equation}
and $\xi\left(\{w_{mi}, \bar{w}_{mi}\}\right)$ is independent of $\mu$. To find $\xi$, we notice that in the limit ${\mu\to \infty}$ the metric~(\ref{LineElementOnM}) behaves as $g_{\mathrm{red}}^{(\mu)}\to \mu\, g_{KE}$, where $g_{KE}$ is the K\"ahler-Einstein metric on $\mathscr{F}$ described in~\eqref{generalKEMetricOnF}. Since its K\"ahler potential is $\sum c_k \log t_k$, from $Ric=g_{KE}$ we find
\begin{equation}
\xi\left(\{w_{mi}, \bar{w}_{mi}\}\right) = |\kappa(\{w_{mi}\})|^2\,e^{- \sum c_k \log t_k} =  |\kappa(\{w_{mi}\})|^2\,\frac{1}{\displaystyle \prod_k t_k^{c_k}}\,,
\end{equation}

\bibliographystyle{utphys}
\bibliography{papers}

\end{document}